\documentclass[aps,prb,twocolumn,floatfix,showpacs]{revtex4}
\usepackage{bm,amssymb,amsmath}

\begin{document}

\title{THE ROLE OF THE EXCHANGE INTERACTION IN THE ONE-DIMENSIONAL
  $n$-COMPONENT HUBBARD MODEL}

\author{E. Szirmai, \"O. Legeza, and J. S\'olyom}

\affiliation{ Research Institute for Solid State Physics and Optics 
\\ H-1525 Budapest, P. O. Box 49, Hungary }

\date{April 30, 2008}

\pacs{71.10.Fd}

\begin{abstract}
  The commensurate $p/q$-filled $n$-component Hubbard chain was investigated
by bosonization and high-precision density-matrix renormalization-group
analysis. It was found that depending on the relation between the number of
components $n$, and the filling parameter $q$, the system shows metallic or
insulating behavior, and for special fillings bond-ordered (dimerized,
trimerized, tetramerized etc.) ground state develops in the insulating
phase. A mean-field analysis shows
that this bond ordering is a direct consequence of the spin-exchange
interaction, which plays a crucial role in the one-parameter Hubbard model ---
not only for infinite Coulomb repulsion, but for intermediate values as well.
\end{abstract}

\maketitle      


Recent experimental results in ultracold gases in optical lattices may be
simulated by multi-component fermionic systems in which spins can take more
than two degrees of freedom \cite{honer}.  A natural candidate for the
description of such systems from the theoretical point of view is the
SU$(n)$-symmetric generalization of the standard SU(2) Hubbard model
\cite{Hubb1-4} which has been investigated intensively in the past by both
analytic and numerical approaches
\cite{honer,marston,assaraf,assaad,szirmai01,LSS,BLSS,rapp,zhao}. In fact this
model may mimic strongly correlated electron systems where the orbital degrees
of freedom of $d$ and $f$ electrons play important role and these extra
degrees of freedom are taken into account by considering $n$-component
fermions.

The Hamiltonian of the model is usually written in the form
\begin{equation} 
  \label{eq:ham}
  {\mathcal H} = \sum_{i=1}^N \bigg[ -t \sum_{\sigma=1}^n (c_{i,\sigma}^\dagger
  c_{i+1,\sigma}^{\phantom \dagger} + c_{i+1,\sigma}^\dagger
  c_{i,\sigma}^{\phantom\dagger}) +
  \frac{U}{2} \sum_{\sigma,\sigma'=1 }^n n_{i, \sigma}n_{i,\sigma'}
  \bigg] ,
\end{equation}
where $N$ is the number of sites in the chain. The operator
$c_{i,\sigma}^\dagger$ ($c_{i,\sigma}^{\phantom \dagger}$) creates
(annihilates) an electron at site $i$ with spin $\sigma$, where the spin index
is allowed to take $n$ different values. $n_{i,\sigma}$ is the particle-number
operator, $t$ is the hopping integral between nearest-neighbor sites, and $U$
is the strength of the on-site Coulomb repulsion.  In what follows $t$ will be
taken as the unit of energy.


It is well known that in the weak-coupling regime the half-filled
$n$-component Hubbard model is an insulator with gapped charge and spin modes
(for $n>2$), while in the large $U$ limit the system can be described --- at
least for even $n$ --- by an effective Heisenberg model \cite{marston,
szirmai01}. Away from half filling, based on leading-order renormalization
group analysis, where the higher-order umklapp processes do not give
contribution, one can find Luttinger liquid behavior with gapless bosonic
charge and spin modes. However, one can expect that these higher order
processes become relevant for commensurate fillings and cannot be
neglected. To see their effect, first we analyze how the spectrum of a
Luttinger liquid is modified by these processes and then we study the
occurrence of phases with spatial inhomogeneity for special fillings.

In order to describe the low temperature physics of the system, one can
consider only that part of the Hilbert space which contains states close to
the Fermi surface. Therefore, in one dimension the dispersion relation can be
linearized around the Fermi points $\pm k_{\mathrm{F}}$. With this
approximation the properties of the fermion problem can be calculated by using
the renormalization group method and bosonization \cite{S-GNT}. One finds that
for generic fillings, the $n$-component Hubbard model --- as well as the usual
$n=2$ component model --- is an $n$-component Luttinger liquid. The well-known
one-particle and particle-hole excitation spectrum of the two-component
Luttinger liquid \cite{H} can be easily generalized for fermions with $n$
internal degrees of freedom:
\begin{equation}
\label{eq:Lutt-Ham-1}
E = \sum_j \hbar u_j \frac{2\pi}{L} \left( n_+^j + n_-^j + \Delta_+^j +
\Delta_-^j \right),
\end{equation}
where the momentum is quantized in units of $2\pi/L$, and $n_\pm ^j$ are
integers describing the particle-hole type bosonic excitations: the term
$j=\mathrm{c}$ describes the excitations of the charge mode which is the
symmetric combination of the bosonic phase fields $\phi_\sigma (x)$ of the
corresponding fermion fields, $\phi_{\mathrm{c}} (x) = \frac{1}{\sqrt{n}}
\sum_{\sigma=1}^n \phi_{\sigma}(x)$. The terms $j=m\mathrm{s}$ describe the
excitations of the $n-1$ spin modes which are independent antisymmetric
combinations of the appropriate boson fields, $\phi_{m\mathrm{s}}(x) =
(m(m+1))^{-1/2}\left[ \sum_{\sigma=1}^m \phi_{\sigma}(x) - m\phi_{m+1}(x)
\right]$. In Eq.~(\ref{eq:Lutt-Ham-1}) $\Delta_\pm^j$ corresponds to the
one-particle excitations: $\Delta_\pm^j = \frac{1}{16} ( \sqrt{K_j} J_j \pm
\delta N_j /\sqrt{K_j} )^2$, where $\delta N_j$ is the change in the number of
particles, and $J_j$ describes the current in the $j$th channel generated by
processes which break the chiral particle-number or spin conservation. Since
the total momentum is given by
\begin{equation}
P=\hbar k_{\mathrm{F}} J_{\mathrm{c}} + \sum_j \hbar \frac{2
\pi}{L}\left(n_+^j - n_-^j + \Delta_+^j - \Delta_-^j \right),
\end{equation}
and the charge current $J_{\mathrm{c}}$ is an even number due to the total
particle-number conservation, in the termodymanic limit soft modes appear not
only at zero momentum but also at even integer multiples of
$k_{\mathrm{F}}=\pi f$ for filling $f$.

For commensurate filling the higher-order umklapp processes are not irrelevant
anymore, and they modify the Luttinger liquid spectrum. In the considered
restricted Hilbert space (with low energy states), the higher-order umklapp
processes can be described only by multiparticle scatterings in fermion
representation, and in these processes the number of scattered particles
depends on the filling factor $f$. Namely, for $f=p/q$ filling, exacly $q$
particles take part in the leading-order multiparticle umklapp
processes. Therefore, due to the Pauli principle these processes are forbidden
for local interaction, if $q>n$, and the system remains Luttinger liquid. If,
on the other hand, $q\leq n$, at least for strong interactions these processes
are relevant, and their contribution has to be taken into account. The
$q$-particle umklapp processes can be described in the terms of the bosonic
fields as
\begin{equation}
\label{eq:umkl}
H_{\mathrm{U}} = g_3 \int \textrm{d}x \sum_{\{\sigma_i\}'}
\cos\big[2\left(\phi_{\sigma_1}(x)+\ldots +\phi_{\sigma_q}(x)\right)\big] \,.
\end{equation}
Here $g_3$ is the coupling of $q$-particle umklapp processes and prime
indicates that the summation over the spin indices gives contribution only for
terms which contains phase fields with different spins. The other terms are
forbidden by the Pauli principle. One can easily see from Eq. (\ref{eq:umkl})
that for $q=n$ the leading-order um\-klapp processes modify only the spectrum
of the symmetric combination of the boson fields, which means, that the
excitation of the charge mode needs a finite energy of order $g_3$, while the
spin modes remain gapless. It is more interesting, as we will see, that if
$q<n$, the leading-order umklapp scatterings couple the spin and charge modes,
and due to this coupling the whole spectrum becomes gapped --- not only at
zero momentum, but at $k^* = 2k_{\mathrm{F}}$, and integer multiples of $k^*$,
too.


Considering the analytical predictions it is expected that the behavior of the
system should be determined by the $k^* = 2k_{\mathrm{F}}$ modes. We
have, therefore, studied the model numerically using the high-precision
density-matrix renormalization group (DMRG) method \cite{white-dmrg} for
several system sizes and values of $q$ and $n$ as a function of $U$.  We have
detected and located quantum phase transition points (QPTs) and determined the
spatial characteristics of the ground state using various quantum information
entropies \cite{vidal-qpt,legeza-qpt,legeza-incomm}
\begin{equation}
s_N(l) = -{\mathrm{Tr}}\big[ \rho_N(l) \ln \rho_N(l) \big] \,,
\label{eq:neumann}
\end{equation}
where a finite block of length $l$ of a long chain of $N$ sites is considered
with the corresponding reduced subsystem density matrix $\rho_N(l)$.  As has
been shown before \cite{vidal-qpt, legeza-qpt}, anomalies in the entropy
functions or in their derivate signal QPTs, and peaks in the Fourier spectrum
of $s_N(l)$ carry information about the position of soft modes (for critical
models) or the spatial inhomogeneity of the ground state (for gapped systems)
\cite{legeza-incomm}.  In the latter case the spatial modulation of the ground
state can be a site- or a bond-centered density wave. A site-centered density
wave would manifest itself in an oscillation of the entropy of single sites or
in the local electron density.  The existence of a bond-centered density wave
can be demonstrated by studying the variation of the bond energy or the
two-site entropy along the chain \cite{legeza-qpt}.

We have found that for models with $q\ge n$, for finite systems, the block
entropy oscillates with a period determined by the filling, but all Fourier
components except for $q=0$ disappear in the $N \rightarrow \infty$ limit.
Therefore, the ground state of the system is spatially uniform.  Using the
analytic form of the block entropy known for critical models \cite{cardy},
\begin{equation}
s_N(l) = \frac{c}{6}\ln \left[ \frac{2N}{\pi} \sin \left( \frac{\pi l}{N}
\right) \right] + g \,,
\label{eq:cardy}
\end{equation}
where $g$ is a shift due to the open boundary, we have also determined the
value of the central charge $c$.
For $q>n$ the model remains critical with $c=n$ for finite $U$ as well
\cite{LSS}, while when $q=n$, a charge gap opens for finite $U>0$, whereas the
spin modes remain gapless, $c=n-1$ \cite{BLSS}, in agreement with the
theoretical expectation. This means that the ground state for $p/n$-filling is
a critical SDW phase without true long-range order.

A significantly different behavior has been found for systems with $q<n$. The
block entropy function saturates beyond some system size \cite{LSS} for finite
$U$ values, indicating that the corresponding models are fully gapped
\cite{vidal-size}.  In fact, gap opens
in the spectrum of all modes for $U>0$ \cite{BLSS}.  Even more interestingly,
the translational symmetry of the Hamiltonian is broken and a spatially
nonuniform ground state emerges whose periodicity depends on the
filling.  Since all Fourier components of the site entropy and local charge
density have been found to vanish for long chains, the ground state is a
bond-centered density wave.  Therefore, we have identified bond-ordered
dimerized, trimerized or tetramerized phases depending on the
filling.



The occurrence of this spatially nonuniform phase was further investigated
analytically on mean-field level in the large $n$ limit, for one-third-filled
system, with $n$ an integer multiple of 3. It was mentioned above that the
change of the periodicity of the translational invariance can be realized by
site- or bond-centered density waves, however, bond ordering cannot be taken
into account within the applied method \cite{LSS} if there is no
nearest-neighbour interaction in the system. In order to analyse the
possibility of bond ordering, too, we have considered a more general model,
the Hubbard-Heisenberg model \cite{marston}, which contains antiferromagnetic
nearest-neighbour spin-exchange $J$, in addition to the on-site Coulomb
repulsion $U$. The Hubbard-Heisenberg Hamiltonian is
\begin{multline}
 {\mathcal H} = \sum_{i=1}^N \bigg[ -t \sum_{\sigma=1}^n (c_{i,\sigma}^\dagger
      c_{i+1,\sigma}^{\phantom \dagger} + H.c. 
      ) \\ + \frac{U}{2} 
      \sum_{\sigma,\sigma'=1}^n n_{i, \sigma}n_{i,\sigma'} + J
      \vec{S}_i \vec{S}_{i+1} \bigg], \, \,
\end{multline}
where we use the same notations as in Eq. (\ref{eq:ham}), and
$\vec{S}_i$ is the SU($n$) spin operator.

We have found that the periodicity of the translational invariance changes for
arbitrary small positive value of $J$: it seems that the Hubbard model is
unstable against the antiferromagnetic nearest-neighbour Heisenberg exchange,
independently of the value of the on-site Coulomb interaction. Considering
this instability and the well-known behaviour of the Hubbard model in the
strong-coupling limit --- where it is equivalent to the Heisenberg model with
an effective antiferromagnetic exchange --- one can conclude, that in the
Hubbard model the Heisenberg interaction becomes relevant in the intermediate
regime, too, not only in the strong-coupling limit. 

The above analysis for the role of the Heisenberg exchange was done only for
$f=1/3$ filling and special values of $n$, therefore the question of the
relevant exchange interaction and its role in the occurrence of the symmetry
broken phase needs further investigation, e.g., it would be interesting to
analyse the phase diagram of the $n$-component $t$-$U$-$J$ model for different
commensurate fillings, and for different values of the couplings. 


This research was supported by the Hungarian Research Fund (OTKA)
Grants No.\ K 68340, F 046356 and NF 61726 and the J\'anos Bolyai Research
Fund.  The authors acknowledge computational support from Dynaflex Ltd. under
Grant No. IgB-32.

\end{document}